\documentclass{article}

\usepackage{multirow}

\usepackage{arxiv}
\usepackage{graphicx}% Include figure files

\usepackage[utf8]{inputenc} % allow utf-8 input
\usepackage[T1]{fontenc}    % use 8-bit T1 fonts
\usepackage[colorlinks=true]{hyperref}       % hyperlinks
\usepackage{url}            % simple URL typesetting
\usepackage{booktabs}       % professional-quality tables
\usepackage{amsfonts}       % blackboard math symbols
\usepackage{nicefrac}       % compact symbols for 1/2, etc.
\usepackage{microtype}      % microtypography
\usepackage{lipsum}
\usepackage{xcolor}
\usepackage{tikz}
\usepackage{makecell}

\usepackage{wrapfig} %to wrap around table

\usepackage{multirow}
\usepackage{tabularx, colortbl}
\usepackage{tabu}
\usepackage{longtable}
\usepackage{array}
\usepackage{booktabs} %nice looking Tables
\usepackage[labelformat=simple, labelsep=colon, justification=justified, labelfont=bf, textfont=normalfont, font=small]{caption}
\newcolumntype{Y}{>{\centering\arraybackslash}X}
\usepackage{caption} 
\usepackage{subcaption}
\usepackage{float}
\usepackage{bm}
\usepackage{enumerate}
\usepackage{dsfont}
\usepackage{pdfpages}
\usepackage{lscape}
\usepackage[subpreambles=true]{standalone}
\usepackage{import}
\usepackage{xr}
\usepackage[super,comma]{natbib}

\usepackage{mathtools}

\title{Prediction of protein allosteric signalling pathways and functional residues through paths of optimised propensity}

\author{
Nan Wu \\
Department of Chemistry \\
Imperial College London \\
\And
Sophia N.~Yaliraki \\
Department of Chemistry \\
Imperial College London \\
\And
Mauricio Barahona\text{*} \\
Department of Mathematics \\
Imperial College London \\
\texttt{m.barahona@imperial.ac.uk} \\
}

\begin{document}
\maketitle

\begin{abstract}
Allostery commonly refers to the mechanism that regulates protein activity through the binding of a molecule at a different, usually distal, site from the orthosteric site. The omnipresence of allosteric regulation in nature and its potential for drug design and screening render the study of allostery invaluable. Nevertheless, challenges remain as few computational methods are available to effectively predict allosteric sites, identify signalling pathways involved in allostery, or to aid with the design of suitable molecules targeting such sites. Recently, bond-to-bond propensity analysis has been shown successful at identifying allosteric sites for a large and diverse group of proteins from knowledge of the orthosteric sites and its ligands alone by using network analysis applied to energy-weighted atomistic protein graphs. To address the identification of signalling pathways, we propose here a method to compute and score paths of optimised propensity that link the orthosteric site with the identified allosteric sites, and identifies crucial residues that contribute to those paths. We showcase the approach with three well-studied allosteric proteins: h-Ras, caspase-1, and 3-phosphoinositide-dependent kinase-1 (PDK1).  Key residues in both orthosteric and allosteric sites were identified and showed agreement with experimental results, and pivotal signalling residues along the pathway were also revealed, thus providing alternative targets for drug design. By using the computed path scores, we were also able to differentiate the activity of different allosteric modulators.
\end{abstract}

\section*{Introduction}

Allostery, first introduced as a term over 50 years ago,~\cite{Monod01011961} is a crucial mechanism in controlling and modulating biological and cellular processes.\cite{Cui2008} 
The classic allosteric effect refers to the modification of the activity at the orthosteric site of an enzyme, due to the binding of a ligand at another, distant site (the allosteric site).
The effect of this distal allosteric binding is transmitted through the structure and dynamics of the biological macromolecule to affect the active site.\cite{He2019} 
The discovery of single-domain allosteric proteins\cite{Volkman2001} and innate dynamic behaviour of enzymes\cite{henziler2007} indicates that allostery is a pervasive phenomenon. All proteins could be allosteric,\cite{Gunasekaran2004} and allosteric regulation of proteins could potentially expand the space of druggable targets.

Controlled up- and down-regulation of protein activity by allosteric modulation is achieved through changes in the conformation and/or dynamics of the protein upon ligand binding events at the allosteric site.~\cite{Kern2003} 
Such induced changes can enhance or reduce the binding affinity of natural substrates at the orthosteric site.\cite{Peracchi2011} Allosteric drugs would thus present advantages, since controlled regulation is hard to attain with traditional orthosteric drugs, which interfere directly with binding at the active site. Furthermore, allosteric targets could also improve drug specificity: the existence of multiple potential allosteric sites on a single protein could provide higher specificity,\cite{Nussinov2013} compared to targeting orthosteric sites, which are well conserved across families of proteins with similar functions.\cite{Christopoulos2004} 
Yet, despite these potential advantages, the underlying mechanisms of allostery are poorly understood\cite{Wodak2019} and a key challenge is to find \emph{bona fide} allosteric sites (or allosteric modifying residues) as drug targets.

Both experimental and computational methods have been developed to identify putative allosteric sites. Experimental methods, including tethering,\cite{HARDY2004706,Erlanson2004} nuclear magnetic resonance (NMR) techniques\cite{Selvaratnam2011,Oyen2013} and high-throughput screening combined with X-ray crystallography,\cite{Rath2000,Wright2002} have successfully found druggable allosteric sites. Yet these methods suffer from laborious screening of huge compound libraries and serendipitous discovery. 
Alternatively, computational methodologies for predicting allosteric sites can take advantage of ever-increasing availability of protein structure data,\cite{Burley2021} including for allosteric proteins (AlloSteric Database (ASD),\cite{Liu2020} ASBench\cite{Huang2015} and CASBench\cite{Zlobin2019} benchmark sets), and rising computational power (for reviews see Refs.\cite{Lu2019, Greener2018}). 
Several computational methods have focused on understanding communication pathways for the identification of allosteric sites using, e.g.,  molecular dynamics (MD) simulations,\cite{Ghosh2007,Shukla2014,Gunsteren2006} measures based on normal mode analysis (NMA) of elastic network models (PARS,\cite{Panjkovich2012,Panjkovich2014} AlloPred\cite{Greener2015} and AllositePro,\cite{Song2017}) as well as machine learning approaches exploiting allosteric site features.\cite{Chen2016,Huang2013} Although these methods have achieved some success in predicting allosteric sites, their prediction accuracy is moderate. Furthermore, only MD simulations keep the atomistic detail of structures, although at the cost of substantial computational power,\cite{Hollingsworth2018} whereas most of the other methods circumvent this limitation through the use of coarse-grained representations of proteins (usually at the level of residues) at the cost of losing resolution.\cite{Collier2013} Different from the general methods discussed above, Guarnera and Berezovsky introduced a structure-based statistical mechanical model of allostery (SBSMMA) relying on calculations of changes in the Gibbs free energy upon perturbation.\cite{Tee2018} Fogha \textit{et al.} analysed the density and clustering of crystallisation additives to find putative allosteric sites.\cite{Fogha2020} Moreover, Wang \textit{et al.} applied a perturbation propagation probability algorithm (Ohm) that acts on an unweighted residue-level graph based on distance-cutoffs to predict hotspots.\cite{Wang2020}

Bond-to-bond propensities\cite{Amor2016} is based on the analysis of high-resolution (atomistic) energy-weighted protein graphs constructed from structural data, and was shown to achieve high prediction accuracy 
%of 95\% on 20 test allosteric proteins 
while staying computationally efficient. 
Recently, the method has been further benchmarked on two large allosteric protein databases, ASBench (118 structures) and CASBench (314 structures), and achieved over 89.8\% and 98.1\% accuracy, respectively.\cite{Mersmann2021,Wu2021} Bond-to-bond propensities quantify how the fluctuations applied to a group of bonds (`source') affect any other bond in the protein, and summarises this impact probabilistically through the \textit{quantile score} (QS), a propensity score computed using quantile regression against the distance to the source. This procedure enables both short- and long-range couplings to be measured, a key feature of allosteric signalling. Importantly, the complexity of protein structures is retained by the fully atomistic graph,\cite{Delmotte2011,Amor2014} yet advances in algorithmic matrix theory\cite{10.1145/2488608.2488724,10.1145/1007352.1007372} make calculation of propensities efficient for large systems. These features make bond-to-bond propensity analysis a cost-effective computational method to predict allosteric sites using high-resolution structures.

It has been argued that allostery can be understood in terms of the propagation of allosteric signals through pre-existing pathways within the protein,\cite{delSol2009} and that understanding such allosteric pathways is essential to determine sites and residues that are strongly coupled to the orthosteric site so as to make allosteric prediction more robust.\cite{Dokholyan2016} Although much effort has been devoted to understanding protein dynamics and allostery via solution NMR,\cite{East2020} our focus here is on available computational methods. Demerdash \textit{et al.} applied machine learning based on structural and network features to predict residues involved in allosteric signalling pathways which then allow the determination of allosteric sites.\cite{Demerdash2009} MCPath uses Monte Carlo to predict potential functional residues and allosteric pathways.\cite{Kaya2013} Botello-Smith \textit{et al.} obtained residue-level protein graphs from MD simulations and useed current-flow betweenness to identify key residues for signal transmission.\cite{doi:10.1021/acs.jctc.8b01197} Ohm, specifically mentions predicted allosteric pathways and important residues identified within the paths, but the method relies on coarse-grained input.\cite{Wang2020} Several of these methods have been validated with experimental results in the literature, but no single method can at once inform functional residues at orthosteric and allosteric sites together with key residues along the allosteric signalling pathways. Moreover, no attempt is made to combine allosteric site prediction and ranking of these sites at the atomistic level. 

The idea of allosteric communication pathways is an intrinsic component of bond-to-bond propensity analysis, as its computation is based on the concept of signals being transmitted from the allosteric to the orthosteric site across the atomistic network. 
Here, we present a method that uses bond-to-bond propensities to compute and assign a quantitative score to signalling pathways that connect the residues of the orthosteric and allosteric sites through paths of optimised propensity. For a full description, see Methods.
Briefly, starting with bond-to-bond propensity analysis on atomistic graphs, all non-covalent interactions in the protein are scored with QS that reflect how strongly the bond is coupled to the `source' (i.e., the orthosteric site). 
Once the atomistic computation of the QS is performed, we add non-covalent interactions, in decreasing order from high to low QS, until the shortest path(s) between the source and a given residue in the allosteric site is found, always meeting the criterion that there is no more than one consecutive step along the backbone. 
This requirement is inspired by the biophysical idea that allosteric signal transmission is akin to anisotropic thermal diffusion, for which heat flow is mainly via non-covalent interactions such that only the surrounding atoms (i.e., the neighbouring residues) are heated.\cite{Ota2005} 
Each \textit{propensity optimised path} (POP) thus obtained is then scored based on its sequence of propensities (see Eq~\eqref{eq:pop_score}), thus allowing us to determine the crucial pairs of residues connecting the orthosteric and allosteric sites. 
In addition, to evaluate key signalling residues along the paths, we consider both residue participation in POPs and the effect that removal from the protein graph of the residue under study has on POP scores and path lengths (i.e., akin to the effect of computational mutagenesis).

We exemplify the method for allosteric signalling pathway detection and identification of key functional and signalling residues on three important enzymes: h-Ras, caspase-1, and 3-phosphoinositide-dependent kinase-1 (PDK1). In h-Ras and caspase-1, we predict correctly allosteric signalling pathways and crucial residues in agreement with experimental mutational studies. For PDK1, not only do we reveal the key signalling pathways and residues but also exploit the pathway scores to compare the effect of different ligands bound to the allosteric site, both qualitatively (inhibition \textit{vs.}\  activation) and quantitatively (level of activity).

\section*{Results}

\subsection*{Uncovering allosteric communication pathways and key signalling residues in h-Ras}

\begin{figure}[ht]
\centering
  \includegraphics[width=15cm]{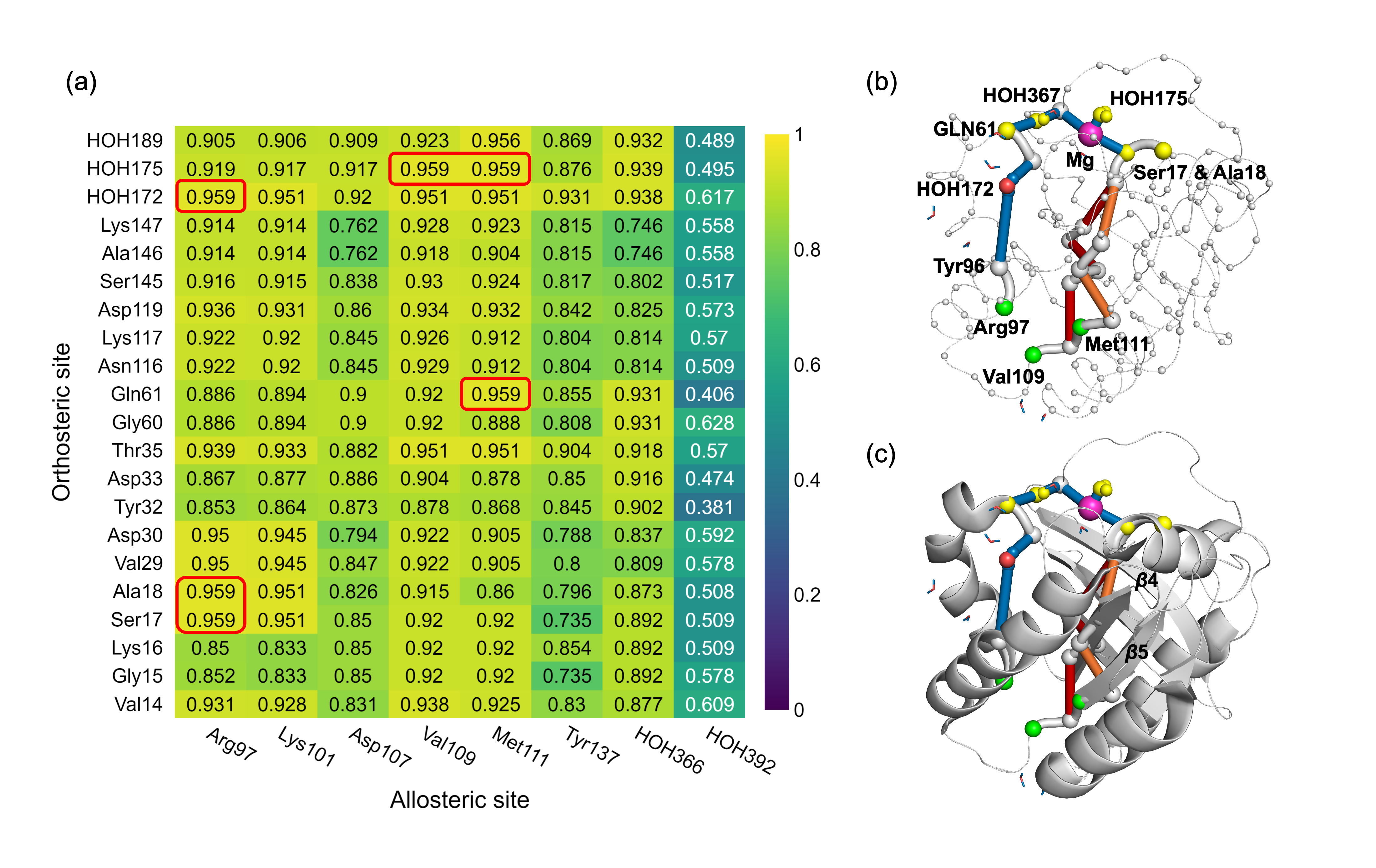}
  \caption{\textbf{Computation of propensity optimised paths between the orthosteric and allosteric site residues for activator-bound h-Ras (PDB ID: 3K8Y).} (a) Scores of propensity optimised paths for all residue pairs, with the highest path score of 0.959 indicated with a red frame. (b) Propensity optimised paths with maximum score of 0.959 for h-Ras. The orthosteric residues (yellow spheres) communicate with allosteric residues (green spheres) via three distinct pathways (coloured in blue, red and orange). In each of the paths, the grey links are steps on the backbone (i.e., involving neighbouring residues) and the coloured connections are non-covalent interactions. (c) Superposition of the three paths onto the structure of h-Ras. The blue path passes through the hydrogen-bond network whereas the other two paths (red, orange) go through the $\beta$4 and $\beta$5 strands.} 
  \label{fig:3K8Y}
\end{figure}

The enzyme h-Ras, also known as transforming protein p21, is a GTPase involved in cell division regulation.\cite{McCormick1995} Calcium acetate has been shown to be an allosteric activator of h-Ras.\cite{Buhrman2010} We started with a bond-to-bond propensity analysis of this structure (PDB ID: 3K8Y) using as `source' its natural substrate, phosphoaminophosphonic acid-guanylate ester (GNP), and  obtained the quantile scores of all non-covalent interactions, including hydrogen bonds, salt bridges and hydrophobic interactions (see also Ref.~\cite{Amor2014}). The orthosteric and allosteric sites were defined as all the residues interacting with, respectively, GNP (the substrate) and calcium acetate (the activator). We then obtained and scored all the propensity optimised paths between all the residues in the orthosteric and allosteric sites. See Methods for a detailed description of the approach. The computed POP scores~\eqref{eq:pop_score} are collated in Figure~\ref{fig:3K8Y}.

The highest path score for h-Ras is 0.959 (see Figure~\ref{fig:3K8Y}a)), and was obtained for paths starting from 5 residues in the orthosteric site (Ser17, Ala18, Gln61, HOH172, HOH175) and reaching only three residues in the allosteric site (Arg97, Val109, Met111). To assess whether these sets of residues have been found to be essential for allosteric signalling, we compared our results to a network of hydrogen bonds connecting the allosteric and orthosteric sites, as proposed by Buhrman \textit{et al}.\cite{Buhrman2010} Within this network, the catalytic (HOH175) and bridging (HOH189) water molecules together with the key catalytic residue Gln61 are essential for the intrinsic hydrolysis in h-Ras, whereas Arg97 and Val109 are responsible for the binding of the acetate ion at the allosteric site. Our results show that four of these five residues are involved in allosteric pathways with the highest path score and HOH189 is involved in the pathway with the second highest score (0.956), which agree with the hydrogen-bonding network. 
Taking a closer look at the POP with the highest score from Ser17, Ala18 and HOH172 to Arg97, we see that they all converge into Gln61, and the final five steps are actually the POP from Gln61 to Arg97 with sequence Gln61-Gly60-HOH367-Tyr96-Arg97 (coloured in blue in Figure~\ref{fig:3K8Y}(b)). This sequence, where three of our propensity optimised paths converge, is indeed a section of the hydrogen-bonding network proposed by Buhrman \textit{et al},\cite{Buhrman2010} further supporting the finding of Gln61 as a key catalytic residue. 
Interestingly, the other three POPs with the highest score (the paths from Gln61 and HOH175 to Met111, and the path from HOH175 to Val109) do not involve the hydrogen-bonding network and reveal a divergent signalling pathway also starting at Gln61. Indeed, these three paths all start with (Gln61)-HOH175-Thr35-Mg168-Ser17-Lys16 and then pass through $\beta$4 and $\beta$5 strands (coloured in red and orange in Figure~\ref{fig:3K8Y}(b)-(c)). The appearance of the water molecule HOH175 and the magnesium ion Mg168 in these additional high score paths highlights their known importance for h-Ras catalytic processes.\cite{Huang1998}

It is illustrative to compare our results to those obtained with other cutting-edge computational methods developed to quantify the coupling between orthosteric and allosteric sites based on other principles. Specifically, we analysed h-Ras with AlloSigMA 2, a recently developed tool based on SBSMMA.\cite{10.1093/nar/gkaa338} By DOWN-mutating the allosteric residues (substituting them by small Ala/GLy-like residues) and quantifying the effect on the orthosteric site using the Allosteric Signalling Map (ASM), Val109 and Met111 are indicated as key allosteric residues, in agreement with our results (Figure~\ref{fig:3K8Y}). 
 On the other hand, ASM does not highlight the importance of the allosteric residue Arg97, which was found to be responsible for the binding of the acetate ion at the allosteric site. In addition, our method explicitly provides the intra-protein signalling pathway linking orthosteric and allosteric residues, thus allowing further understanding of structural information and access to other relevant residues and interactions that could be targeted. Detailed information can be found in Supplementary Information Part 1 and Figure S1.

It is also worth noting that when analysing h-Ras without the activator calcium acetate (PDB ID: 3LBN), we find substantially altered high-scoring POPs connecting the orthosteric and allosteric sites.
In particular, the top six POPs for 3LBN, which connect HOH202, Lys16 and Ser17 in the orthosteric site to Val109 and Lys101 in the allosteric site, traverse the Switch II region, as well as the $\alpha 1$ and $\beta 1$, $\beta 2$ and $\beta 3$ strands.  Hence, some of the most prominent allosteric signalling paths in h-Ras are only activated upon binding of the activator ligand.  Yet, key paths involving Gln61 still include water molecules (HOH227 in 3LBN), thus underscoring their crucial role in h-Ras allostery (detailed paths and scores for 3LBN can be found in Supplementary Information Part 2 and Figure S2).

Returning to h-Ras with the bound activator (PDB ID: 3K8Y), we then identified key residues along the computed allosteric signalling paths by computing two measures of residue importance (see Methods). First, we obtained the residue participation frequency in the computed POPs. In total, there are 153 residues in the protein (excluding the orthosteric and allosteric sites) and 68 (out of 153) are involved in the POPs connecting the 21 residues in the orthosteric site with the 8 residues in the allosteric site.
On average, each of those 68 residues participates in 10 POPs but some residues have much higher participation frequency in those paths. The top five residues according to their POP participation frequency are shown in Table \ref{table:3K8Y_res_frequency}. 
Tyr96, Ile100 and HOH367 are the most involved residues in allosteric signalling pathways. Of these, HOH367 and Tyr96 lie on the shortest path from Gln61 to the allosteric site within the hydrogen-bonding network,\cite{Buhrman2010} and Ile100 forms a hydrogen bond with Tyr96 of high propensity (QS = 0.91), indicating the importance of Tyr96. 
The other two residues do not feature in the hydrogen-bonding network and might hint at other unknown pathways.

\begin{table}[ht]
\caption{\textbf{Top five signalling residues within the POPs (excluding orthosteric and allosteric sites), according to frequency of participation in the POPs computed for h-Ras (PDB ID: 3K8Y).}}
    \renewcommand{\arraystretch}{}
    \begin{center}
    \begin{tabularx}{0.5\textwidth}{|Y|Y|}
    \hline
    \textbf{Residue} & Residue frequency\\
    \hline
    Tyr96 & 46\\
    \hline
    Ile100 & 43\\
    \hline
    HOH367 & 42\\
    \hline
    Leu133 & 34\\
    \hline
    Ile84 & 29\\
    \hline
    \end{tabularx}
    \end{center}
    \renewcommand{\arraystretch}{1}
\label{table:3K8Y_res_frequency}
\end{table}

As a second measure of residue importance, we computed the impact that the computational removal of a residue has on pathway scores and lengths. To do this, we mimic  \textit{alanine scanning mutagenesis} computationally by removing from the atomistic graph all the weak interactions associated with each residue of h-Ras, one residue at a time. This approach has been shown to be an efficient and robust computational means to mimicking the effect of alanisation, under the assumption that no large structural rearrangements occur.\cite{Amor2014,Peach2018} The POP scores and lengths were then recomputed in the modified graph, and we obtained z-scores for each residue (see Methods). Table \ref{table:3K8Y_z_score} shows the top five residues with the highest z-scores with respect to path scores and path lengths.
The magnesium ion (Mg168), which is known to be a key to facilitating GNP binding,\cite{Huang1998} appears as the most influential according to both scores. The z-scores of the POP scores are low and do not uncover any other residues in the hydrogen bond network. This is not surprising since POPs involve bonds with high QS, and the removal on one residue only reroutes the paths through other residues with high QS, with little change in the POP scores. Therefore, only key bonds essential in maintaining a high POP score, in this case the electrostatic interactions with Mg ion, are found.
The z-scores for path lengths, on the other hand, reveal more structural features and indicate residues whose removal leads to longer paths. The bonds and local structure associated with such residues are responsible for connecting two sites with the shortest paths. Again, we find Mg168 again as the top residue, followed by HOH367 and Tyr96, consistent with the hydrogen bond network in~Ref.~\cite{Buhrman2010}.

\begin{table}[ht]
\caption{\textbf{Top five signalling residues within the POPs (excluding orthosteric and allosteric sites), according to z-scores computed from the change in path score and path length upon residue removal for h-Ras (PDB ID: 3K8Y).}}
\centering
    \renewcommand{\arraystretch}{}
    \begin{tabularx}{.7\textwidth}{|Y|Y||Y|Y|}
    \hline
    \textbf{Residue} & z-score \qquad (path score) & \textbf{Residue} & z-score \qquad (path length)\\
    \hline
    Mg168 & 11.13 & Mg168 & 8.22\\
    \hline
    Val112 & 1.55 & HOH367 & 5.18\\
    \hline
    Ser106 & 1.39 & Tyr96 & 5.08\\
    \hline
    Ser136 & 1.36 & Ser136 & 2.41\\
    \hline
    Glu62 & 1.22 & Thr58 & 1.46\\
    \hline
    \end{tabularx}
    \renewcommand{\arraystretch}{1}
\label{table:3K8Y_z_score}
\end{table}

\subsection*{Key allosteric residues in caspase-1}

\begin{figure}[ht]
\centering
  \includegraphics[width=11cm]{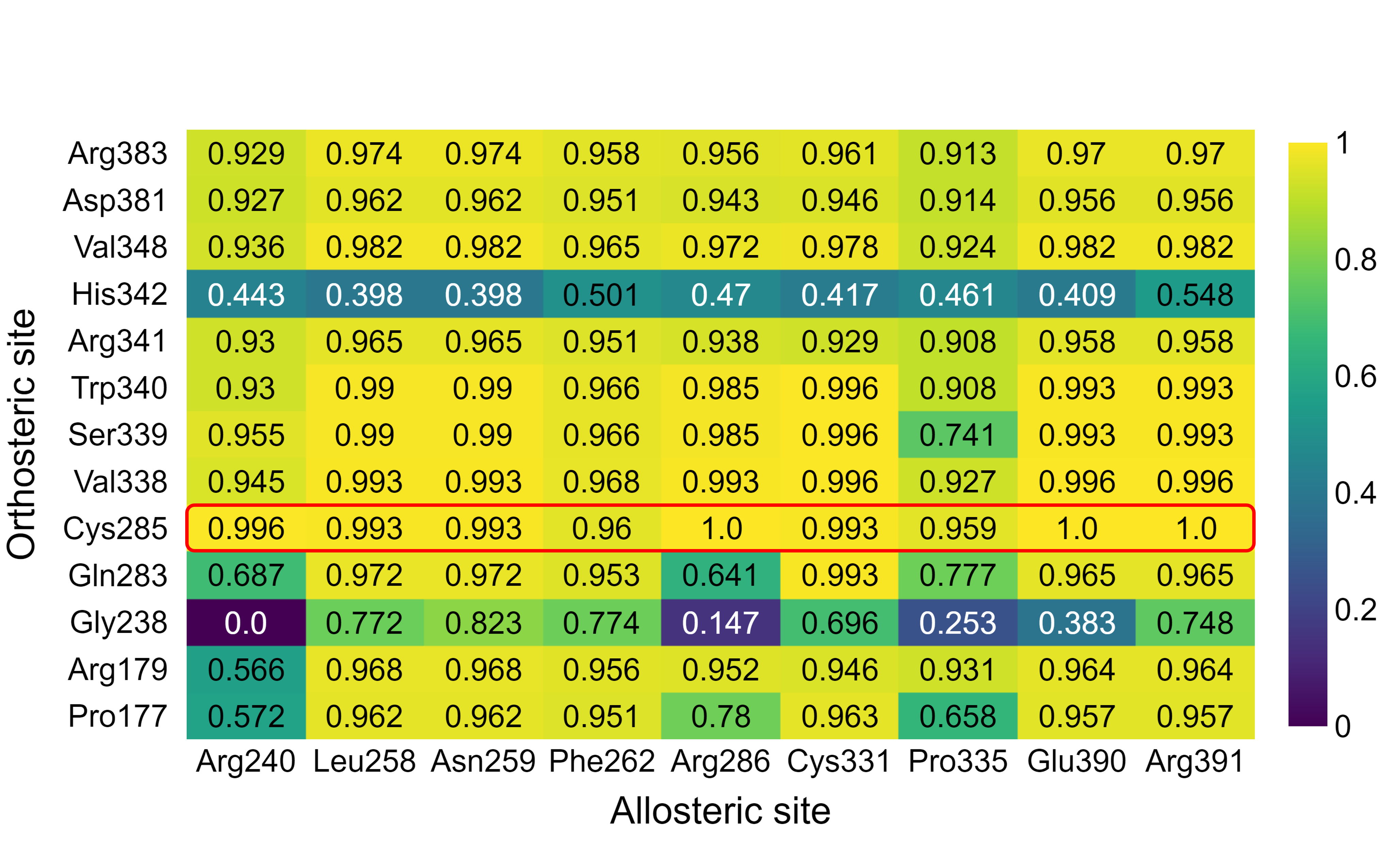}
  \caption{\textbf{Scores of propensity optimised paths between the residues of the orthosteric and allosteric sites for wild type caspase-1 (PDB ID: 2HBQ).} Highlighted (red frame) is the strong connection from Cys285, a key residue at the orthosteric site, to the allosteric residues.} 
  \label{fig:2HBQ}
\end{figure}

Caspase-1 is a well-studied allosteric protein, which plays an important role in apoptotic processes in the cell.~\cite{Wilson1994} In Ref.~\cite{Amor2014}, bond-to-bond propensity was used to uncover the allosteric site using the crystal structure (PDB ID: 2HBQ) of wild-type human caspase-1 in complex with the orthosteric ligand z-VAD-FMK~\footnote{z-VAD-FMK corresponds to the inhibitor carbobenzoxy-valyl-alanyl-aspartyl-[O-methyl]- fluoromethylketone}, 
Here, we take this analysis further to find out functionally important residues in connecting the orthosteric and allosteric sites. 
Following the procedure described above, we found and scored propensity optimised paths for each residue pair between the orthosteric and allosteric sites. These scores are compiled in Figure~\ref{fig:2HBQ}.  
Note that caspase-1 is a tetramer formed by two identical heterodimers, and the propensity calculation is indeed conducted on the complete tetramer, but we show the path scores are presented for half of the structure (chains A and B)  since the scores for the other half (chains C and D) are the same.

Figure~\ref{fig:2HBQ} reveals a strong connection between the orthosteric site (formed in this case by 13 residues) and the allosteric site (9 residues), as indicated by the high POP scores. Cys285, to which z-VAD-FMK is covalently attached, is the residue in the orthosteric site that is most strongly to the allosteric site residues (with most path scores $\geq$ 0.99). Residues Val338, Ser339, and Trp340 are also strongly coupled to the allosteric site. 
Focusing on the residues of the allosteric site, we found that Arg286, Glu390 and Arg391 connect strongly to Cys285 (with path scores equal to 1.0), noting that the path from Cys285 to Arg391 contains Arg286 and Glu390 as intermediate residues. Hence, Arg286 and Glu390 are crucial in the orthosteric-allosteric coupling, in agreement with experimental results that show them to be part of a hydrogen bond network connecting orthosteric and allosteric sites, and, additionally, the fact that the interaction between these two residues is disrupted upon allosteric inhibitor binding.\cite{Datta2008} 

To explore further the importance of these two residues,
we repeated the analysis for the structures of two experimentally resolved caspase-1 mutants: Arg286Ala (PDB ID: 2HBR) and Glu390Ala PDB ID: 2HBY). The results are presented in Supplementary Information Part 3 and Figure S3.
Upon Arg286Ala and Glu390Ala mutations, the path scores from Cys285 to Arg286 remain unchanged at 1.0, whereas the POP score from Cys285 to Glu390 drops from 1.0 (the top allosteric signalling pathway in wild-type caspase-1, 1/117 paths) to 0.88 and 0.937, respectively, rendering this pathway out of the top 10 signalling pathways for the mutants (78/117 paths for Arg286Ala  and 48/117 paths for Glu390Ala). This indicates that
Arg286 and Glu 390 are both key signalling residues that induce the loss of the strong orthosteric-allosteric connection in the observed experimental structures.

We also computed our measures of residue importance (residue frequency, and the two z-scores) for the wild type caspase-1 (PDB ID: 2HBQ),  as described above. The residues with top 10 residue frequencies in all POPs were: Asn337, Ile282, His237, Asn263, Ser332, Asp336, Phe330, Ala329, Cys244 and Thr180. 
The results for both z-scores are shown in Table \ref{table:2HBQ_z_score}.
Experimental evidence has shown that the Ser332Ala mutant has a 3.7-fold reduction in catalytic efficiency\cite{Datta2008} and this residue is identified as the residue with fifth highest path participation frequency and the third highest z-score on the disruption of path lengths. 
In addition, the Asn337Ala mutant has a $\sim$2-fold reduction in catalytic efficiency\cite{Datta2008} and we found that this residue has the top path participation frequency and the sixth highest z-score of path scores. 
To the best of our knowledge, none of the other identified residues have been studied experimentally in the literature.

\begin{table}[ht]
\caption{\textbf{Top 10 key signalling residues within all allosteric pathways according to z-scores based on change in path score and path length for wild type caspase-1 (PDB ID: 2HBQ).}}
    \renewcommand{\arraystretch}{}
    \centering
     \begin{tabularx}{.7\textwidth}{|Y|Y||Y|Y|}
    \hline
    \textbf{Residue} & z-score \qquad (path score) & \textbf{Residue} & z-score \qquad (path length)\\
    \hline
    Ser289 & 12.52 & Asp336 & 10.71\\
    \hline
    Ile282 & 3.71 & Ser289 & 6.57\\
    \hline
    Cys244 & 3.40 & Ser332 & 3.91\\
    \hline
    Ile243 & 3.04 & Ser136 & 3.03\\
    \hline
    Ile239 & 2.86 & Thr58 & 2.73\\
    \hline
    Asn337 & 2.38 & Asp108 & 2.44\\
    \hline
    Ser236 & 2.24 & His94 & 2.44\\
    \hline
    His237 & 2.01 & Gln129 & 2.14\\
    \hline
    Gly287 & 1.72 & Glu126 & 2.14\\
    \hline
    Phe349 & 1.71 & Val125 & 1.85\\
    \hline
    \end{tabularx}
    \renewcommand{\arraystretch}{1}
\label{table:2HBQ_z_score}
\end{table}

To make our measures more specifically targeted on the high scoring POPs, we considered paths with POP score equal to and above 0.99.  The residue pairs involved in such high scoring POPs, shown in Table~\ref{table:2HBQ_qs_path_res},
are  Asn337 (six times), Ser332 (4 times), Asp336 (once) and Ala284 (once). Of these, Asn337 and Ser332 have appeared as important residues above, and have been shown experimentally to have a lower catalytic efficiency upon alanisation.\cite{Datta2008} Therefore, considering high-scoring pathways and their corresponding intermediate residues can provide further insight for the determination of key signalling residues. 

\begin{table}[ht]
\caption{\textbf{Orthosteric and allosteric residues involved in paths with score $\geq$ 0.99 for wild type caspase-1 (PDB ID: 2HBQ).}}
    \renewcommand{\arraystretch}{}
    \centering
    \begin{tabularx}{.7\textwidth}{|Y|Y|}
    \hline
    Orthosteric residue(s) & Allosteric residue(s)\\
    \hline
    Cys285 & Val124, Val127, Arg131, Phe147, Thr148\\
    \hline
    Val338, Ser339, Trp340 & Leu258, Asn259, Arg286, Cys331, Glu390, Arg391\\
    \hline
    \end{tabularx}
    \renewcommand{\arraystretch}{1}
\label{table:2HBQ_qs_path_res}
\end{table}

\subsection*{Allosteric communication pathways in PDK1 under various allosteric modulators}

In the previous examples, we have shown that computing and scoring propensity optimised paths allows us to identify crucial orthosteric, allosteric and signalling residues involved in allosteric signalling pathways. We now investigate whether the paths and scores can be used to understand and quantify the allosteric activity of different allosteric ligands.

To pursue this line of enquiry, we considered 3-phosphoinositide-dependent kinase-1 (PDK1), a member of the AGC kinase family.\cite{MORA2004161} 
PDK1 is constitutively active owing to the auto-phosphorylation at Ser241 (Sep241 upon phosphorylation), and it plays a critical role in phosphorylating and activating more than 23 AGC related kinases. It is also indicated in tumourigenesis.\cite{Raimondi2011} Indeed, PDK1 is important in regulating the intracellular PI3K-AKT pathway, where its activation is closely related to human cancers.\cite{Hennessy2005} Therefore, PDK1 modulators are attractive as anti-cancer agents. 

There are three ligand binding sites on PDK1: the catalytic ATP binding site, the substrate binding site, and the PDK1 Interacting Fragment (PIF) binding site.\cite{Schulze2016} Targeting the ATP binding site (orthosteric site) with competitive inhibitors generally leads to low selectivity, as this site is conserved in over 500 protein kinases.\cite{BOGOYEVITCH2007622} The PIF pocket is not only employed as a binding site for downstream substrate kinases, but also exploited for stimulation of PDK1 activity.\cite{Schulze2016} Hence, the PIF pocket behaves as an allosteric site and has spawned the development of non-competitive PDK1 modulators. An  important example is PIFtide, a 24-amino acid polypeptide derived from PDK1 substrate PRK2, which binds to the PIF pocket and enhances 7-fold the activity of PDK1.\cite{emboj.7601416} 

\begin{figure}[ht]
\centering
  \includegraphics[width=14cm]{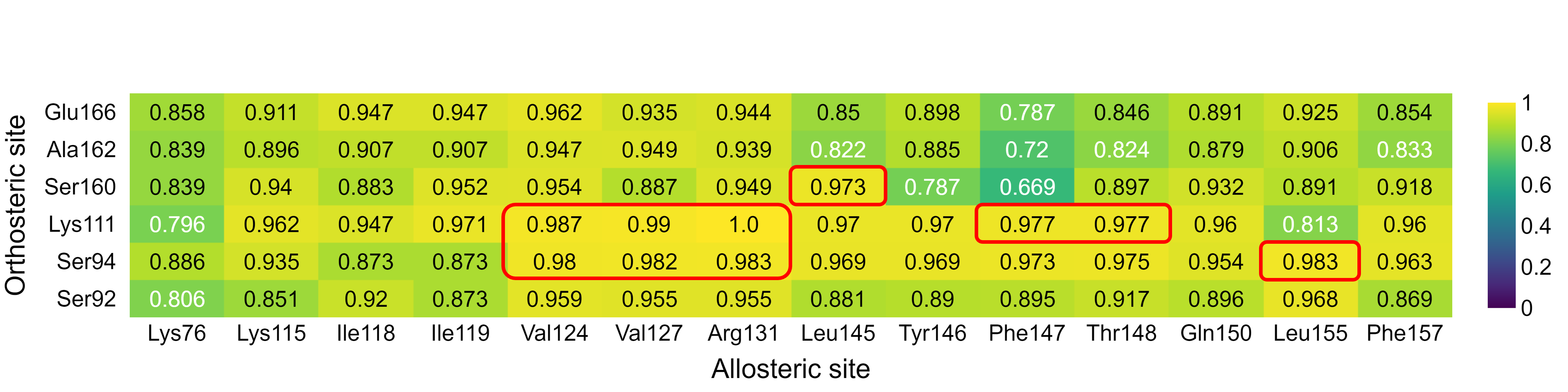}
  \caption{
  \textbf{Scores of propensity optimised paths between the residues of the orthosteric and allosteric sites for PIFtide bound PDK1 (PDB ID: 4RRV) with top 10 path scores highlighted.}} 
  \label{fig:4RRV}
\end{figure}

Here, we used the crystal structure of PDK1 complexed with PIFtide (PDB ID: 4RRV, with ATP removed) for propensity analysis, where we defined the orthosteric site as the residues responsible for ATP binding, and the allosteric site as all the residues interacting with PIFtide. 
The results of our propensity optimised path computations are summarised in Figure \ref{fig:4RRV}, where we also highlight the the top 10 scoring POPs.

The key functional residues in the orthosteric and allosteric sites involved in the high-scoring POPs can be seen in Figure \ref{fig:4RRV}.
The crystal structure solved by Rettenmaier \textit{et al.} illustrates how the allosteric signal can be transmitted to the ATP binding site.\cite{Rettenmaier2014} PIFtide binds to Arg131 which results in the movement of the helix $\alpha$C and leads to a change in Glu130. The ensuing hydrogen bonding between Glu130 and Lys111 then conducts the allosteric signal to the ATP binding site. Reassuringly, the highest scoring path (1.0) in our analysis was precisely Arg131–Glu130–Lys111, which is in agreement with the experimental results in Ref.~\cite{Rettenmaier2014} 
Alanine scanning mutagenesis conducted experimentally by Rettenmaier \textit{et al.} determined binding energy hotspots on PIFtide and highlighted five residues (Phe14, Asp16, Phe17, Asp18, Tyr19) as the most contributing residues.\cite{Rettenmaier2014} 
We found that Arg131 interacts with four of those five residues, whereas Thr148, Gln150 and Leu155 interact with two of them. This indicates that the key allosteric residues Arg131, Thr148 and Leu155, which were found amongst the paths with highest propensity (Table~\ref{table:4RRV_qs_path_res}), match the experimental evidence, thus confirming their role in activation signalling.

\begin{table}[ht]
\caption{\textbf{Orthosteric and allosteric residues involved in the top 10 allosteric signalling paths predicted with path score $\geq$ 0.99 for PIFtide bound PDK1 (PDB ID: 4RRV).}}
    \renewcommand{\arraystretch}{}
    \centering
    \begin{tabularx}{0.7\textwidth}{|Y|Y|}
    \hline
    Orthosteric residue(s) & Allosteric residue(s)\\
    \hline
    Ser94 & Val124, Val127, Arg131, Phe147, Thr148\\
    \hline
    Lys111 & Val124, Val127, Arg131, Leu155, Thr148\\
    \hline
    \end{tabularx}
    \renewcommand{\arraystretch}{1}
\label{table:4RRV_qs_path_res}
\end{table}

\begin{figure}[ht]
\centering
  \includegraphics[width=14cm]{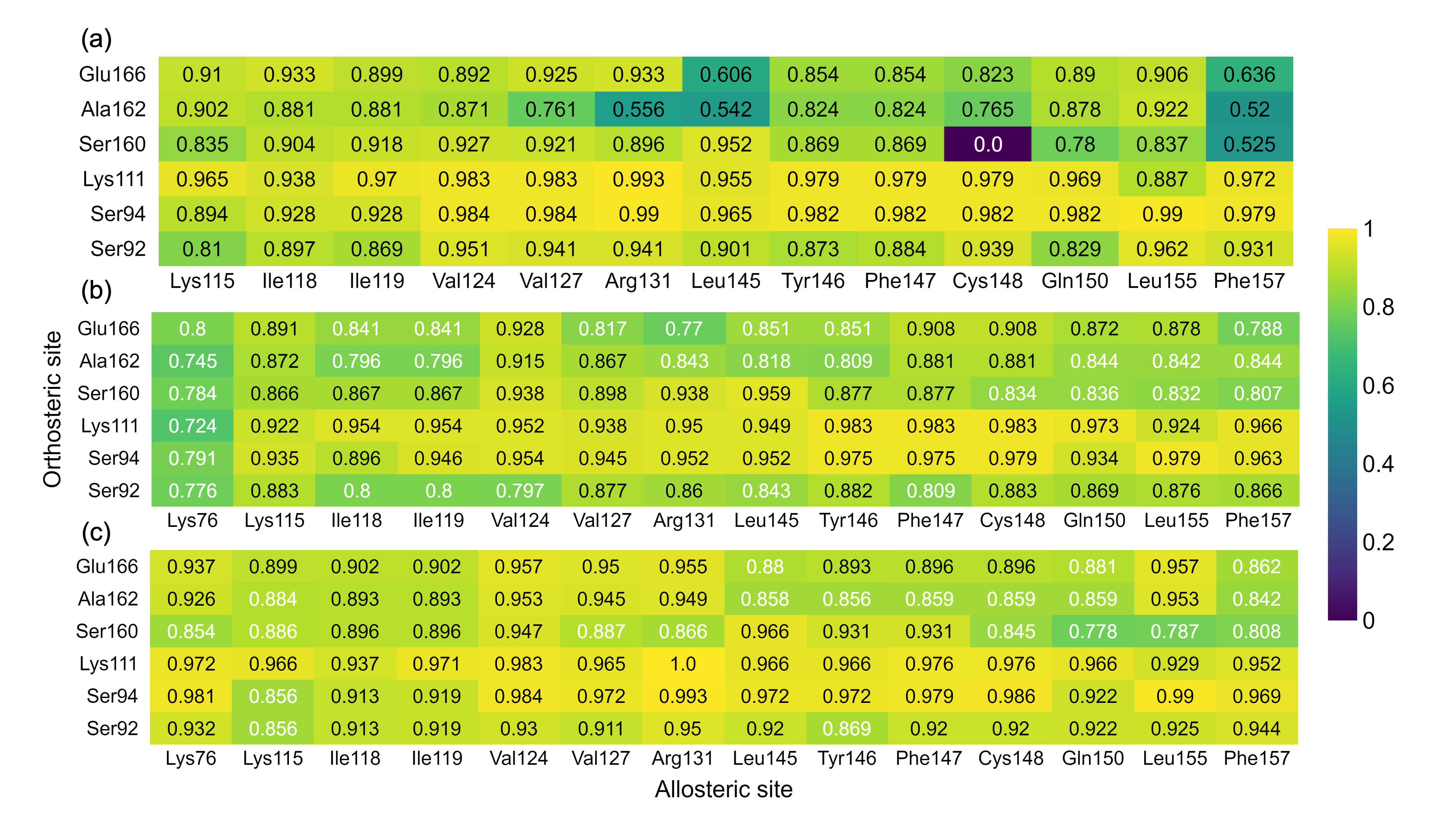}
  \caption{
   \textbf{Scores of propensity optimised paths between the residues of the orthosteric and allosteric sites for
   three PDK1 structures bound to different activators and inhibitors with top 10 path scores highlighted.} (a) Strong activator (J30) bound PDK1 (PDB ID: 3OTU). (b) Weak activator (2A2) bound PDK1 (PDB ID: 3ORZ). (c) Inhibitor (1F8) bound PDK1 (PDB ID: 3ORX). Note that Lys76 does not appear in the PDB structure 3OTU.
   } 
  \label{fig:PDK1_ligands}
\end{figure}

To further test this hypothesis, we analysed the structures of PDK1 complexed with three different allosteric ligands:  a strong activator J30 (PDB ID: 3OTU), a weak activator 2A2 (PDB ID: 3ORZ), and an inhibitor 1F8 (PDB ID: 3ORX). Using the same orthosteric and allosteric residues for the PDK1-PIFtide complex (except that Lys76 is not present in 3OTU), all the paths were computed and scored for each of the structures and the results are shown in Figure \ref{fig:PDK1_ligands}.

Comparing the scores for PDK1 bound to the strong activator J30 (Figure \ref{fig:PDK1_ligands}(a)) to the scores obtained for the PIFtide bound PDK1 (Figure \ref{fig:4RRV}), we found broad agreement: seven of the top ten scoring paths are the same and the key residues identified are generally the same. Indeed, both J30 and PIFtide are strong allosteric activators of PDK1 and the POP results confirm this. 
Figure \ref{fig:PDK1_ligands}(b) shows the results for the weak activator 2A2. We found that although the top ten scoring POPs are different, the key residues highlighted are similar to those found with strong activators.
On the other hand, when the inhibitor 1F8 is bound, the patterns of the signalling paths are strongly modified, as seen in Figure \ref{fig:PDK1_ligands}(c). Note how all the top ten scoring paths concentrate in one region, and the allosteric residues Lys76, Val124, Val127 and Arg131 do not appear as functionally important. 

\begin{figure}[ht]
\centering
  \includegraphics[width=12cm]{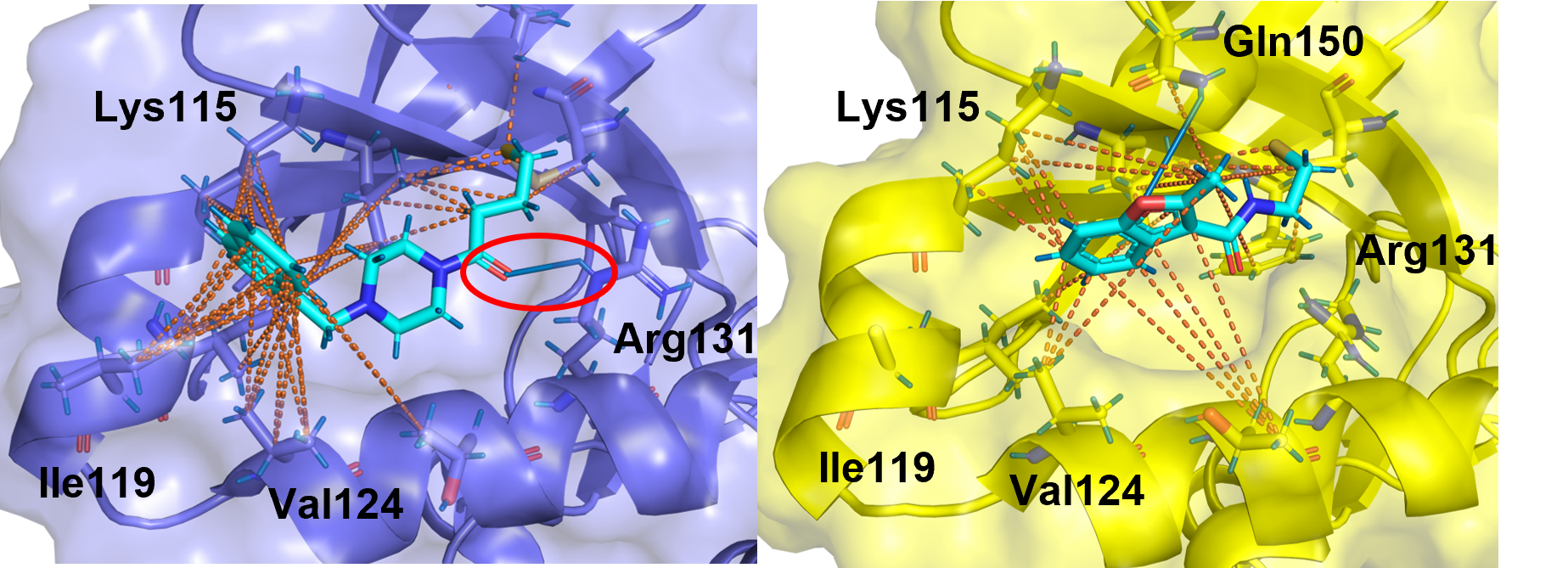}
  \caption{\textbf{Details on ligands–PIF pocket interactions.} (a) Non-covalent interactions formed between the strong allosteric activator J30 with the PIF pocket. The orange dotted lines are hydrophobic interactions identified with BagPype.~\cite{florian_song_2022_6326081} (b) Non-covalent interactions formed between the allosteric inhibitor 1F8 with the PIF pocket. Lys115, Val124 and Gln150 are for positioning of the ligand and no hydrogen bond is observed with Arg131.} 
  \label{fig:Lig_compare}
\end{figure}

Combining the analyses of the structures bound to activators (J30, 2A2) and to the inhibitor (1F8) with the key allosteric residues in the PDK1-PIFtide structure, it emerges that interaction with Arg131 is essential for PDK1 activation, 
since only the inhibitor 1F8 does not make any interaction with Arg131. More specifically,   
Figure~\ref{fig:Lig_compare} shows that J30 forms a hydrogen bond with Arg131, whereas 1F8 does not interact with Arg131 at all. This structural analysis suggests that Lys115, Ile119, Val124 and Gln150 only play a role in positioning the ligand, but it is the hydrogen bond (circled in red) formed between the amide oxygen atom in J30 and the nitrogen bound hydrogen atom in Arg131 that initiates the allosteric signalling. 
Few inhibitors (other than 1F8) have been discovered for PDK1. An example  is the alkaloid derivatives discovered by Bobkova \textit{et al.}, which showed selectivity towards PDK1 and inhibition. Although no crystal structure was obtained for those compounds, \textit{in silico} docking indicated that no binding with Arg131 is present, owing to a lack of a carboxyl moiety.\cite{Bobkova2010} This again supports the argument that interaction with Arg131 could be indispensable for PDK1 activation, and avoidance of this interaction through careful design of molecules could result in an allosteric inhibitor.

The previous results show that our path analysis in PDK1 allows us to identify a qualitative difference between activating and inhibiting allosteric ligands. Our computations can also be used to quantify the effect of different ligands, and we present here some initial work in this direction. As discussed above, PDK1 remains active owing to the auto-phosphorylation of Ser241 into Sep241, and allosteric signals can also be transmitted from the allosteric site through Sep241 to control the protein activity.\cite{Rettenmaier2014} Therefore, we examined the connection of the allosteric site to Sep241, which specifically keeps PDK1 active.

To investigate this issue, we calculated the POP scores from the modulator-bound allosteric residues to Sep241 for 11 PDB structures of PDK1 bound with different allosteric modulators (mostly activators). These include the 4 structures analysed above in Figures~\ref{fig:4RRV}~and~\ref{fig:PDK1_ligands} together with 7 additional experimentally resolved structures. The average POP scores from allosteric residues to Sep241 for all these structures are summarised in Table \ref{table:PDK1_ligands}.
We found that the average POP score from the allosteric site to Sep241 is highest (0.97) for 1F8, the only inhibitor in the set, whereas all the activators have scores lower than 0.86. Within the activators, we found that the POP scores are ordered as follows:
\begin{itemize}
    \item  Table~\ref{table:PDK1_ligands}: PS210 > RS2 > J30 > PIFtide > RS1 $\gg$ PS182 > PS171 > PS114 > PS48 $\gg$ 2A2
\end{itemize}
which compare well with experimental results measuring the activating strength of the ligands: 
\begin{itemize}
    \item Ref.~\cite{Rettenmaier2014}: PS210 > RS2 > PIFtide > RS1
    \item Ref.~\cite{BUSSCHOTS20121152}: PS210 > PS182 > PS48
    \item Ref.~\cite{Sadowsky2011}: J30 > 2A2
    \item Ref.~\cite{LOPEZGARCIA20111463}: PS114 > PS171
\end{itemize}
The activating strengths of the allosteric activators are all correctly ranked based on the average path scores, except for PS114 and PS171. However, we note that PS114 is not completely resolved in the PDB structure (4A06), which could explain the mismatch with experimental data. 

Furthermore, the detailed structural information of the POPs reveals differences between the inhibitor and the activators. In particular, the inhibitor (3ORX) has high POP scores in two specific pathways (i.e., those connecting Val127 and Leu155 to Sep241), whereas the allosteric activators have low POP scores for those pathways. This and other distinct patterns (see Figure S4) can provide further information to potentially distinguish the relevant communication pathways for inhibitor and activators.
Although a fuller comparison is not possible due to the lack of available experimental data, our results show promise to link the structural patterns and magnitude of POP scores with the type and strength of the allosteric modulation. For details of the POPs scores of the PDK1 set, see Supplementary Information Part 4 and Figure S4.

\begin{table}[ht]
\caption{\textbf{Details of 11 PDK1 allosteric modulators.} The modulators are ordered according to the average path scores from the allosteric site residues to Sep241 for 11 allosteric ligands bound to PDK1 structures.}
    \renewcommand{\arraystretch}{}
    \centering
    \begin{tabularx}{.7\textwidth}{|Y|Y|Y|Y|}
    \hline
    PDB ID & Allosteric ligand & Ligand type & Average path score to Sep241\\
    \hline
    3ORX & 1F8 & Inhibitor & 0.97\\
    \hline
    4AW1 & PS210 & Activator & 0.86\\
    \hline
    4RQV & RS2 & Activator & 0.85\\
    \hline
    3OTU & J30 & Activator & 0.84\\
    \hline
    4RRV & PIFtide & Activator & 0.83\\
    \hline
    4RQK & RS1 & Activator & 0.81\\
    \hline
    4AW0 & PS182 & Activator & 0.77\\
    \hline
    4A07 & PS171 & Activator & 0.75\\
    \hline
    4A06 & PS114 & Activator & 0.73\\
    \hline
    3HRF & PS48 & Activator & 0.71\\
    \hline
    3ORZ & 2A2 & Activator & 0.60\\
    \hline
    \end{tabularx}
    \renewcommand{\arraystretch}{1}
\label{table:PDK1_ligands}
\end{table}

\section*{Discussion}

Using protein structural data and applying bond-to-bond propensity analysis to atomistic protein graphs, we are able to quantify how strongly coupled a bond is to the selected source through a propensity quantile score. 
Adding bonds of increasing propensity quantile score to a simplified residue graph, we compute and score propensity optimised paths between the residues of the orthosteric and allosteric sites. The path scores allow us to identify important residues at both ends of the path. We can also find key intermediate (signalling) residues along the path using the residue participation frequency in the paths as well as measures to evaluate the impact of residue removal in path score and length.

We have demonstrated this computational approach with three allosteric proteins: h-Ras, caspase-1 and PDK1. Analysis of h-Ras correctly finds the key catalytic residue Gln61, and identifies Mg168, HOH367 and Tyr96 as the crucial signalling residues based on POP participation frequencies and z-scores of residue removal. These findings are consistent with a hydrogen bond network proposed in the literature.\cite{Buhrman2010} Regarding caspase-1, focusing on the covalently bound Cys285 as the starting point of allosteric signalling pathways, we find that the highest path scores are to Arg286 and Glu390, two functionally important allosteric residues, and subsequent pathway analysis on Arg286Ala and Glu390Ala mutants demonstrates the loss of connection between the orthosteric site and Glu390 upon mutations. The analysis of high scoring POPs highlights the crucial signalling residues Asn337 and Ser332 which, together with Arg286 and Glu390, are in agreement with the hydrogen bond network tested experimentally. Lastly, in PDK1, we find the correct allosteric signalling pathway, Arg131–Glu130–Lys111, and suggest that interaction with Arg131 as the essential element for allosteric activation, which also allows to distinguish activators and inhibitors. We also ranked allosteric activators of PDK1 by scoring the paths from the allosteric site to the key activating residue Sep241, and found good agreement with experimental results. In summary, the analysis introduced here constructs short paths of high propensity based on bond-to-bond propensities to identify functionally important orthosteric, allosteric and signalling residues, and was applied to distinguish allosteric activators from inhibitors and to rank allosteric modulators. 

Further ongoing work includes the robust computation of multiple pathways to allow for a probabilistic evaluation of paths between `source' and `target'. 
Beyond strict optimality, the distribution of scores of an ensemble of optimised paths would provide complementary insight into the robustness of the allosteric signalling between the residue pairs.
In this respect, concepts such as effective distance,\cite{doi:10.1126/science.1245200} residue betweenness and centrality\cite{doi:10.1021/acs.jctc.8b01197} could be interesting for the optimisation and scoring of multiple pathways. 
We also note that bond-to-bond propensity analysis itself has been shown to be able to predict allosteric sites. Hence, combining both analyses, we can first predict the allosteric site, rank the sites based on pathway scores and then identify the key binding residues. The key residues could then act as the starting point for the design of allosteric drug molecules that specifically interact with the residues highlighted. Since the residues would be distributed within the predicted allosteric site, small fragments bound to such residues can be selected first and pieced together to form the final drug molecule, along the lines of fragment-based drug design.\cite{10.3389/fchem.2020.00093} Moreover, our approach could be combined with ligand-site interaction data to guide effective molecule design for a given target on a protein with the aid of supervised machine learning. 
Through validation and testing against data on pathways and functional residues from different allosteric proteins, the approach could be further optimised and developed with the aim to guide the analysis of unknown allosteric mechanisms for proteins and the identification of new structural targets.

\section*{Materials and methods}

\subsection*{Protein structure and sites definition}

The structures of caspase-1, h-Ras and PDK1 in PDB format shown in Table~\ref{table:PDB_structures} were retrieved from the RCSB Protein Data Bank (PDB),\cite{Burley2021} and further processed to remove irrelevant solvent molecules for crystallisation and ions such as chloride ions. The orthosteric and allosteric sites are defined as the interacting residues with the corresponding ligands based on protein graphs constructed in Section~\textit{Construction of the atomistic protein graph}.

\begin{table}[ht]
\caption{\textbf{Details of protein structures used.}}
    \renewcommand{\arraystretch}{1.5}
    \begin{tabularx}{\textwidth}{|Y|Y|Y|}
    \hline
    Protein & Description & PDB ID\\
    \hline
    \multirow{3}{*}{Caspase-1} & Wild type & 2HBQ\cite{Scheer2006} \\ \cline{2-3}
    & Arg286Ala & 2HBR\cite{Scheer2006} \\ \cline{2-3}
    & Glu390Ala & 2HBY\cite{Scheer2006} \\
    \hline
    \multirow{2}{*}{H-Ras} & Wild type, with calcium acetate & 3K8Y\cite{Buhrman2010}\\ \cline{2-3}
    & Wild type, without calcium acetate & 3LBN\cite{Buhrman2010}\\
    \hline
    \multirow{2}{*}{PDK1} & Allosteric activator & 3HRF,\cite{Hindie2009} 3ORZ, 3OTU,\cite{Sadowsky2011} 4A06, 4A07,\cite{LOPEZGARCIA20111463} 4AW0, 4AW1,\cite{BUSSCHOTS20121152} 4RQK, 4RQV, 4RRV\cite{Rettenmaier2014} \\ \cline{2-3}
    & Allosteric inhibitor & 3ORX\cite{Sadowsky2011} \\
    \hline
    \end{tabularx}
    \renewcommand{\arraystretch}{1}
\label{table:PDB_structures}
\end{table}

\subsection*{Construction of the atomistic protein graph}

An energy-weighted atomistic graph is constructed from the 3-dimensional coordinates of the atoms of a protein. The atoms are represented as nodes and the bonds and interactions (covalent and non-covalent) as edges. Edges are weighted based on the interaction energies calculated with relevant interatomic potentials: covalent bonds by bond-dissociation energies;\cite{Huheey1993} hydrogen bonds and salt bridges by the modified Mayo potential;\cite{Dahiyat1997,osti_5941560} hydrophobic interactions by a hydrophobic potential of mean force;\cite{Lin2007} electrostatic interactions by the OPLS (optimized potentials for liquid simulations) potential functions.\cite{doi:10.1021/acs.jctc.8b01197} Details can be found in Refs.\cite{Amor2016,Delmotte2011,Amor2014}. 

To construct the atomistic protein graphs, we used the software package  \underline{B}iochemical \underline{a}tomistic \underline{g}raph construction in \underline{Py}thon for \underline{p}rot\underline{e}ins (BagPype).\cite{florian_song_2022_6326081} The process starts from the cartesian coordinates of atoms from a PDB file. Missing hydrogen atoms are added using Reduce.\cite{Word1999} Covalent bonds are identified using standard covalent bond length cutoffs. All weak interactions are detected and weighted based on the above-mentioned potentials with a 9 Å distance cutoff for hydrophobic interactions, and a 0.01 kcal/mol energy cutoff for hydrogen bonds. The constructed energy-weighted atomistic graph is shown in Figure \ref{fig:Path_computation}.

\subsection*{Bond-to-bond propensity analysis}

Bond-to-bond propensity analysis was introduced and extended in Refs.\cite{Amor2016,Hodges2018}. The analysis quantifies how a perturbation at the bonds of the source is transmitted to all other bonds of the protein via a graph-theoretical measure defined in edge-space. 
Consider a graph with $n$ nodes and $m$ edges.
The $m \times m$ edge-to-edge transfer matrix $M$ was first introduced in Ref.~\cite{Schaub2014} to examine nonlocal edge-coupling in graphs, and was then utilised to analyse atomistic protein graphs in Ref.~\cite{Amor2016}. $M$ can be  interpreted as a Green function (see Refs.~\cite{Schaub2014,Amor2016}), such that the effect of a perturbation at edge $i$ on edge $j$ is described by the element $M_{ji}$ of the matrix 
\begin{equation}
    M=\frac{1}{2}WB^TL^\dag B.
\end{equation}
Here $B$ is the standard $n \times m$ incidence matrix linking the $n$ nodes and $m$ edges; $W$ = diag($w_{ij}$) is an $m$ × $m$ diagonal matrix containing all the edge interaction energies; and the $n \times n$ matrix $L^\dag$  is the pseudo-inverse of the energy-weighted Laplacian matrix $L$, which governs the diffusion dynamics on the graph.\cite{biggs_1974,Lambiotte2014RandomWM}

We can then define the bond propensity, which is the effect of a perturbation from bonds $b^{'}$ (within the source), belonging to the source, on bond $b$ of the protein:
\begin{equation}
    \Pi_{b}^{\text{raw}}=\sum_{b^{'} \in \text{source}} \left|M_{bb^{'}}\right|
\end{equation}
The raw propensity is normalised by the sum of all the bond propensities of the protein:
\begin{equation}
    {\Pi}_{b}=\dfrac{\Pi_{b}^\text{raw}}{\sum_{b} \Pi_{b}^\text{raw}}
\end{equation}
For full details of the derivations, see Ref.~\cite{Amor2016}.

\subsection*{Quantile regression (QR) and computation of the quantile scores}

The propensity of a bond decreases naturally as the distance of the bond to the source increases. Hence, we aim to detect bonds with significantly large bond-to-bond propensity relative to other bonds at a similar distance from the source. To achieve this, we apply conditional quantile regression (QR)\cite{Koenker2001} with distance as the independent variable. 
Specifically, let $d$ be the minimum Euclidean distance from a point in space to any of the bonds of the source residues, calculated based on the midpoint of the bonds. Given that the bond-to-bond propensities, $\Pi_{b}$, decay exponentially with the distance from the source, the following QR minimisation problem is solved for a given quantile $p$ to obtain the estimated parameters of the model:
\begin{equation}
\label{eq:model_QR}
%    \hat{\bm{\beta}}^\text{prot}(p):=
    \left(\hat{\beta}_{0}^\text{prot}(p),\hat{\beta}_{1}^\text{prot}(p)\right) =
    \underset{(\beta_{0},\beta_{1})}{\arg\min}
    \sum_{b}^{\text{protein}}\rho_{p}\left(\log(\Pi_{b})-(\beta_{0}+\beta_{1}d)\right),
\end{equation}
% to obtain the estimated parameters of the model:
% \begin{equation}
% \label{eq:model_QR}
% \hat{\bm{\beta}}^\text{prot}(p)=(\hat{\beta}_{0}^\text{prot}(p),\hat{\beta}_{1}^\text{prot}(p)). 
% \end{equation}
where $\rho_{p}(\cdot)$ is the tilted absolute value function:
\begin{equation}
    \rho_{p}(y)=\left|y\left(p-\mathds{1}(y<0)\right)\right|,
\end{equation}
and $\mathds{1}(\cdot)$ is the indicator function.
% This optimisation gives us our regression model described by the set of  parameters 
% \begin{equation}
% \label{eq:model_QR}
% \hat{\bm{\beta}}^\text{prot}(p)=(\hat{\beta}_{0}^\text{prot}(p),\hat{\beta}_{1}^\text{prot}(p)). 
% \end{equation}

Let $d_{b}$ be the distance between bond $b$ and the source, i.e., the minimum Euclidean distance between $b$ and any of the bonds of the source residues. From the estimated model for the protein~\eqref{eq:model_QR},
the quantile score (QS) of bond $b$ with propensity $\Pi_{b}$ at distance $d_{b}$ is quantile $p_{b}$ given by:
\begin{equation}
\label{eq:QS}
    p_{b}=\underset{p \in [0,1]}{\arg\min}
    \left|\log(\Pi_{b})-
    \left(\hat{\beta}_{0}^\text{prot}(p)+\hat{\beta}_{1}^\text{prot}(p) \, d_{b} \right)
    \right|
\end{equation}
Clearly, the QS of all the bonds in the protein is  in the range between 0 and 1. The higher the QS of a bond, the stronger the bond couples to the source bonds. For more details see Ref.~\cite{Amor2016}

\subsection*{Computation of propensity optimised paths between residue pairs}

Based on the energy-weighted atomistic protein graph described above and the QS~\eqref{eq:QS} obtained through bond-to-bond propensity analysis for all weak interactions, the graph is simplified to the residue level by considering each residue as a node and creating weighted edges between residues by selecting the bond with the highest QS between any residue pair and using the QS as its weight. The backbone of the protein is maintained by adding an edge between two neighbouring residues. To compute the pathway between two residues, edges are added in order of decreasing QS. Starting from the top 5\%, we attempt to find the shortest path between any given residue pair with the constraint of allowing no more than one consecutive step along the protein backbone. This allows us to restrain the path through the weak interactions within the protein and still take adjacent residues into consideration. A suitable biophysical analogy for this process is the transmission of allosteric signals as anisotropic thermal diffusion, for which heat flow is mainly via non-covalent interactions since only the surrounding atoms are heated, hence including the neighbouring residues.\cite{Ota2005} If no path is found based on this criterion, the next 5\% of the edges (of decreasing QS) is added, and we repeat the protocol until a path is found. 
The path is then scored by taking the geometric mean of the QS of all weak interactions $b$ in the path, which we term the propensity optimised path (POP) score:
\begin{equation}
\label{eq:pop_score}
    \Sigma_\text{POP} = \sqrt[k]{p_{b_{1}} p_{b_{2}} \cdots p_{b_{k}}},
\end{equation}
where the POP has $k$ weak interactions $b_i,\ i=1,\ldots,k$.
If there are multiple paths with the same length, the one with the highest path score is selected as the POP for the residue pair. Note that there is no QS associated with a backbone hop; hence backbone hops are thus ignored in the calculation of the path score~\eqref{eq:pop_score}. 

\begin{figure}[ht]
\centering
  \includegraphics[width=16cm]{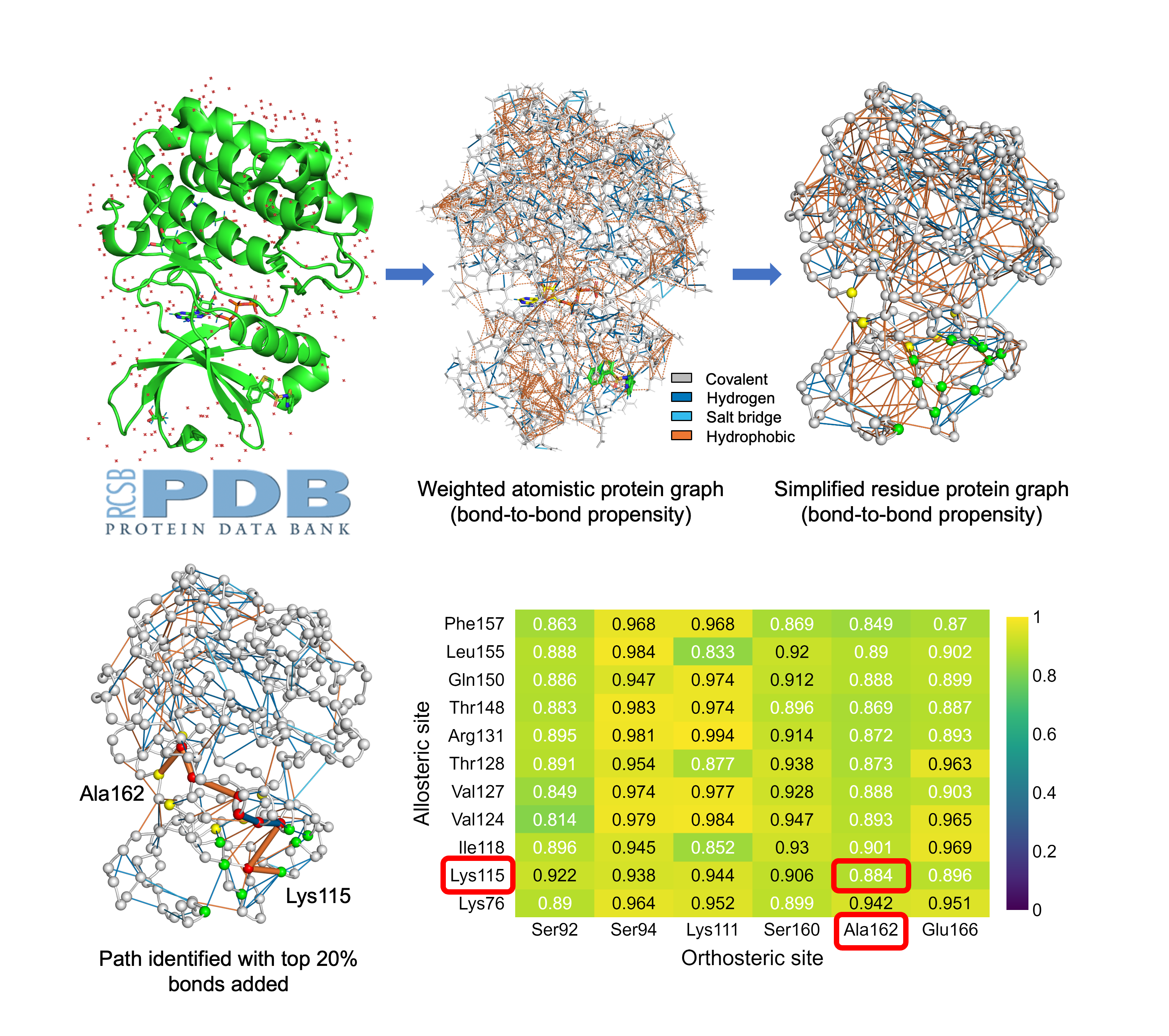}
  \caption{\textbf{Workflow of pathway computation using PDK1 (PDK ID: 4RQK)\cite{Rettenmaier2014} as an example.} The X-ray crystallographic protein structure in PDB format is retrieved and processed as described in Section \textit{Protein structure and sites definition}. An energy-weighted atomistic protein graph is constructed with weak bonds weighted applying bond-to-bond propensity analysis. The atomistic protein graph is then simplified to the residue level to keep the highest scoring bonds between any residue pair. The orthosteric site residues are coloured in yellow and allosteric site residues in green. Five percent of the bonds are added at a time starting from bonds with QS = 1.0 until a path meeting the criteria is identified. For instance, 20\% of the bonds were added in order to find a path between Ala162 and Lys115. The computed path was highlighted in bold with the intermediate residues in red. The paths of each residue pair between the orthosteric and allosteric sites were computed and summarised in a heatmap.} 
  \label{fig:Path_computation}
\end{figure}

Figure \ref{fig:Path_computation} illustrates the full workflow of computing the pathways starting from the initial structural data.

\subsection*{Determination of key signalling residues}

To understand which residues are crucial for allosteric signalling in the paths, we employ two measures. 

\textit{Residues with high participation in propensity optimised paths:} The most direct measure is to count the number of appearances of each residue in all the orthosteric-allosteric POPs (i.e, participation residue frequencies). The residues that appear more often are considered important for allosteric signalling.

\textit{Residues with high influence in pathway scores and length upon removal:} Another measure, adapted from del Sol \textit{et al.},\cite{delSol2006} is the change in score and length of the propensity optimised paths under residue removal. This workflow is repeated by removing the weak interactions of each residue in the protein and computing the change in both the average path length and score of all the computed propensity optimised paths between the orthosteric and allosteric sites:
\begin{equation}
    \Delta X_{r}=\left|X - X_{r, rem}\right|
\end{equation}
where $X$ is the original average path length (or score) and $X_{r, rem}$ is the average path length (or score) when weak interactions involving residue $r$ are removed. Finally, a z-score is used to evaluate the importance of residue $r$.
\begin{equation}
    z_{r} = \frac{\Delta X_{r} - \overline{\Delta X}}{\sigma}
\end{equation}

\section*{Data availability}
All data presented in this study are available at figshare with DOI: \href{https://doi.org/10.6084/m9.figshare.20182166}{10.6084/m9.figshare.20182166}.

\section*{Acknowledgements}

We acknowledge helpful discussions with Florian Song and L\'{e}onie Str\"{o}mich. This work was funded by a President’s PhD Scholarship, Imperial College London, to N.W. Funding from EPSRC award EP/N014529/1 supporting the EPSRC Centre for Mathematics of Precision Healthcare at Imperial is gratefully acknowledged by S.N.Y.and M.B.

\section*{Author contributions}
N.W., M.B. and S.N.Y. conceived the study. N.W. performed the computations and created the figures. All authors analysed the data and wrote the manuscript.

\section*{Competing interests}
The authors declare no competing interests.

\section*{Materials \& Correspondence}
Requests for data and code to m.barahona@imperial.ac.uk.

\bibliographystyle{naturemag}
\bibliography{manuscript}



\section*{Supplementary material}

# Supplementary Information

## 1. Allosteric free energy evaluation of h-Ras (PDB ID: 3K8Y) using AlloSigMA 2

We used AlloSigMA 2[1] to study h-Ras. We applied Allosteric Signalling Map (ASM) to investigate the key binding residues in the allosteric site of 3K8Y (Job ID: ESKO5KG2). We mutated each of the allosteric site residues and noted the $\Delta g$ of the orthosteric site residues for DOWN-mutations to quantify the effect of the allosteric residues. The results are summarised in Fig.S1. We chose DOWN-mutations since the allosteric site here serves to activate h-Ras, hence any destabilising effect would indicate that the residue is a crucial allosteric residue.

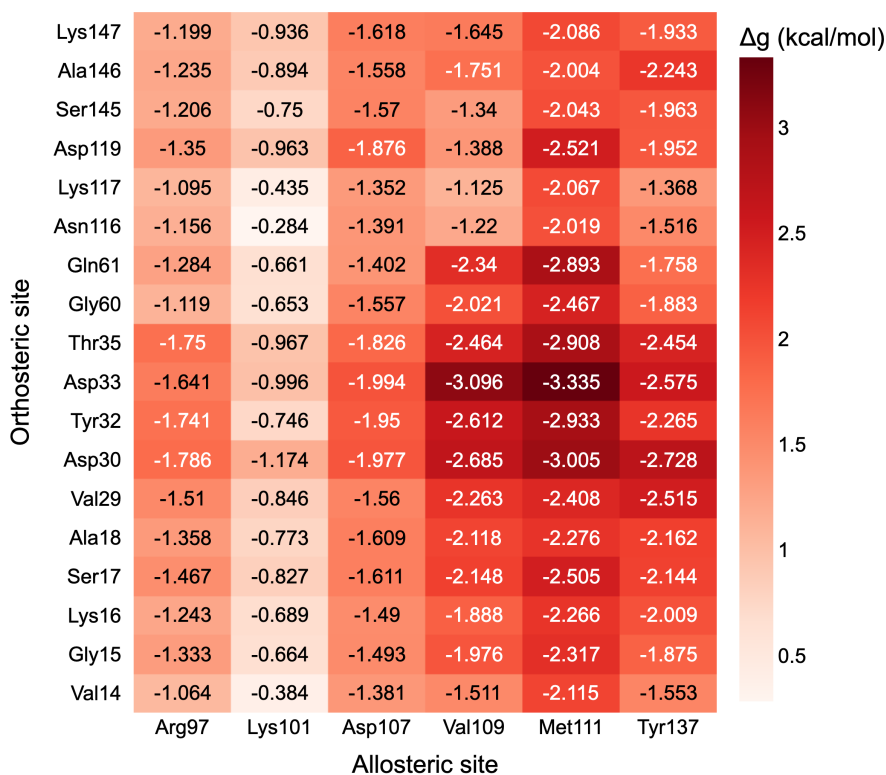

Figure S1: **Part of the Allosteric Signalling Map (ASM) summarising the allosteric effect of allosteric residues on the orthosteric site for h-Ras (PDB ID:3K8Y).**

## 2. Propensity Optimised Path Analysis of h-Ras with no allosteric activator (PDB ID: 3LBN)

We have applied our POP analysis to h-Ras without an allosteric activator (PDB ID: 3LBN). As shown in Fig.S2, the top six paths for this structure connect HOH202, Lys16 and Ser17 in the orthosteric site to Val109 and Lys101 in the allosteric site. These pathways are distinct from those computed for h-Ras with the activator (PDB ID: 3K8Y) reported in Figure 1 in the main text. Hence, we find that some of the most prominent allosteric signalling paths in h-Ras are only activated upon binding of the activator allosteric ligands. Interestingly, the path connecting Gln61 to Arg97 (Gln61-Gly60-HOH367-Tyr96-Arg97), which is key in 3K8Y, has an equivalent path in 3LBN (Gln61-Glu62-HOH227-Asp92-Ile93-Arg97) with a diminished score of 0.865 but also involving the key water molecule HOH227 at the same position in the path. This reinforces the crucial role of water molecules in connecting the orthosteric and allosteric sites.

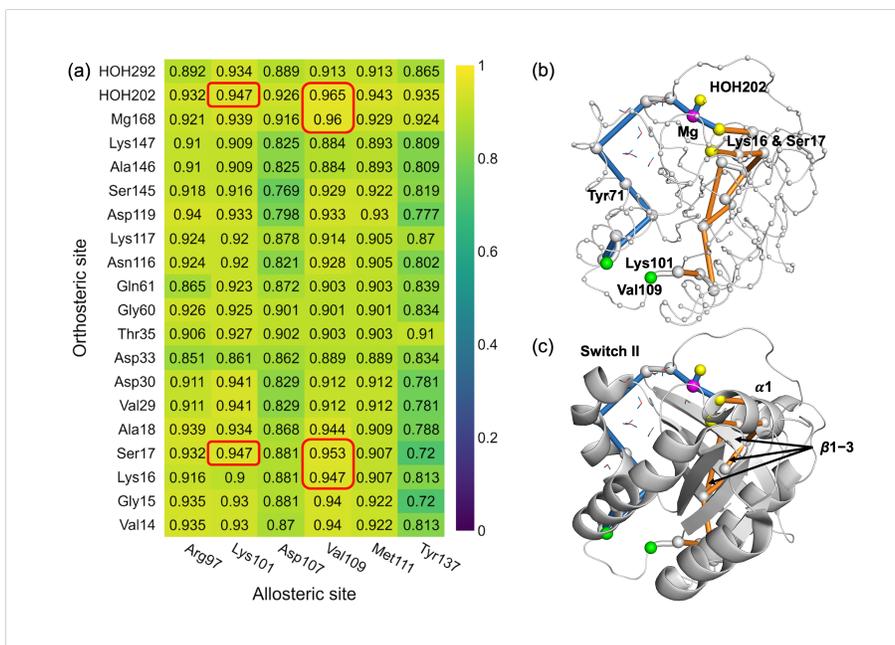

Figure S2: **Computation of propensity optimised paths between the orthosteric and allosteric site residues for h-Ras without the allosteric activator (PDB ID: 3LBN).** (a) Scores of propensity optimised paths for all residue pairs, with the top six scoring paths indicated with a red frame. (b) Top six propensity optimised paths for h-Ras without an allosteric activator (PDB ID: 3LBN). The orthosteric residues (yellow spheres) communicate with allosteric residues (green spheres) via two distinct pathways (coloured in blue and orange). The blue path connecting Ser17 and HOH202 to Lys101 scores 0.947. The orange path connecting Lys16, Ser17, Mg168 and HOH202 to Val109 includes paths with scores of 0.947, 0.953, 0.96 and 0.965 respectively. (c) Superposition of the three paths onto the structure of h-Ras. The blue path goes through the Switch II region while the orange path passes $\alpha 1$ and $\beta 1$, $\beta 2$ and $\beta 3$ strands.

## 3. Propensity Optimised Path Analysis of caspase-1 mutants (Arg286Ala & Glu390Ala)

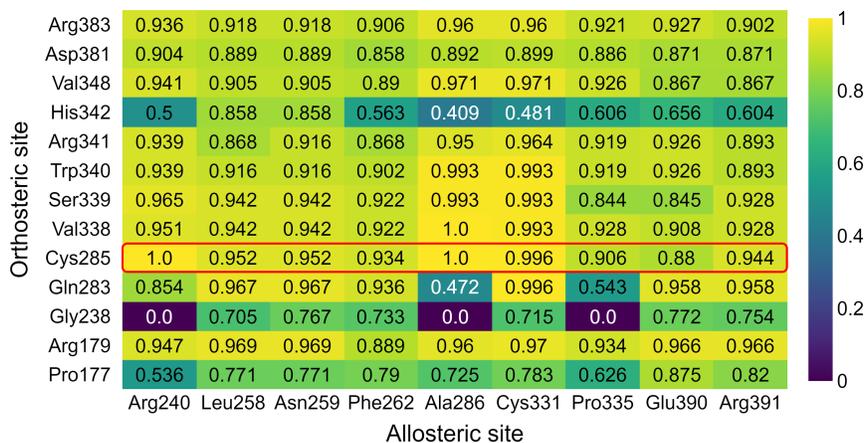

(a) Caspase-1 mutant Arg286Ala (PDB ID: 2HBR).

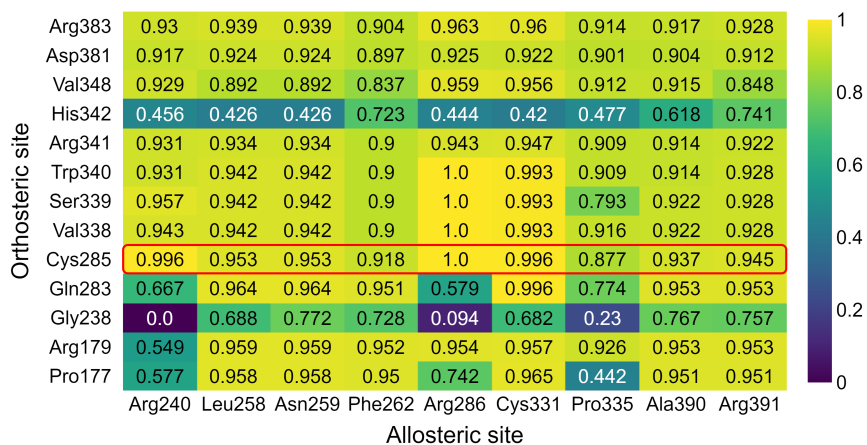

(b) Caspase-1 mutant Glu390Ala (PDB ID: 2HBY).

Figure S3: **Scores of propensity optimised paths between the residues of the orthosteric and allosteric sites for two mutants of caspase-1.** Highlighted (red frame) is the connection from Cys285 (a key residue at the orthosteric site) to all allosteric residues.

3

## 4. Propensity Optimised Path Analysis of PDK1 with different allosteric modulators: inhibitor *vs.* activators

We present further details of our POP analysis of 11 PDK1 structures with different modulators. The scores of POPs connecting the modulator-interacting residues in the allosteric site to the key residue Sep241 are summarised in FigureS4. As discussed in the text, the average POP score for the PDK1 inhibitor (3ORX) is the highest (0.97), whereas the scores of the activators are all below 0.86. Furthermore, we are able to use their scores to rank the activators in agreement with experiments.

Furthermore, our POP calculations provide detailed structural information of the optimised pathways that reveal differences between the inhibitor and the activators. In particular, we observe that the inhibitor (3ORX) has high POP scores in two specific pathways, i.e., those connecting Val127 and Leu155 to Sep241. In contrast, all the allosteric activators have low POP scores for those specific pathways. This and other distinct patterns in FigureS4 provide further information to distinguish the allosteric inhibitor from the activators, and could provide further avenues to investigate important interactions.

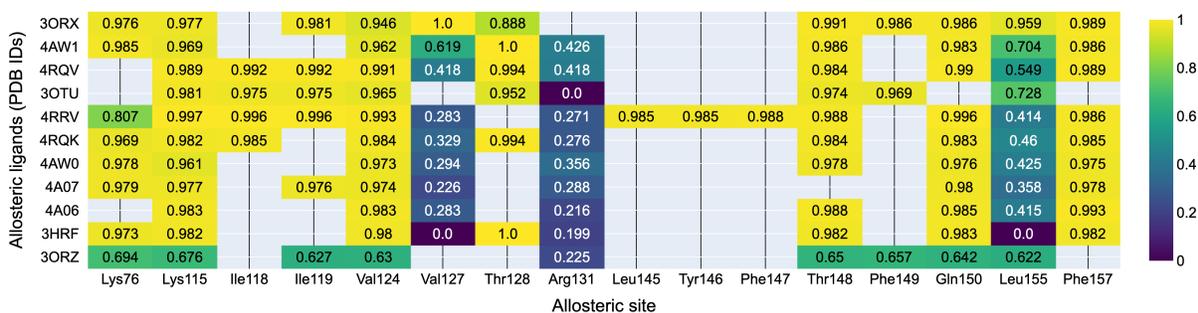

Figure S4: **Scores of propensity optimised paths between the allosteric residues and Sep241 for 11 PDK1 structures.** The residue pairs with no POP score mean that the allosteric residues do not interact with the allosteric modulators.

4



\end{document}